\documentclass[12pt]{article}
\usepackage{graphicx}
\usepackage{amsmath}
\usepackage{cite}
\usepackage{hyperref}
\usepackage{svg}
\usepackage{algorithm}
\usepackage{algpseudocode}

\title{ALCMean's: Unsupervised Community detection using local Laplacian, automatic detection of the number of centers}
\author{Shahin Momenzadeh, Rojiar Pir Mohammadiani\thanks{Corresponding author: Department of Computer Engineering, University of Kurdistan, Sanandaj, Iran. Email: \href{mailto:email@example.com}{r.pirmohamadiani@uok.ac.ir‏}}}
\date{}

\begin{document}
	
	\maketitle
	
	\begin{abstract}
		Community detection is a fundamental problem in the analysis of complex networks. It has applications across social, biological, and financial domains. Traditional algorithms, such as Louvain, LPA, and modularity optimization often require manual parameter tuning. They also suffer from inaccurate cluster center selection and struggle with scalability. To address these challenges, we propose Automatic Laplacian Centrality Means (ALCMean's), a novel community detection algorithm. AlCMean's combines Laplacian energy–based automatic center identification with DeepWalk embeddings for robust node representation. Unlike existing Laplacian-based and clustering methods, AlCMean's eliminates the need to predefine the number of communities, enhances cluster center selection using structural importance, and leverages representation learning for more accurate and stable assignments. Experimental results on benchmark datasets demonstrate 10–20\% higher NMI and ARI scores compared to Louvain, Newman–Girvan, LPA, Fast-Greedy, and a recent GNN-based competitor (MAGI, KDD’24). Additional evaluations with modularity and F1-scores confirm the superiority of AlCMean's. Ablation studies highlight the critical contributions of each component. Despite its reliance on DeepWalk parameters and increased runtime relative to lightweight heuristics, AlCMean's consistently outperforms state-of-the-art methods. This makes it a promising tool for real-world network analysis. The source code and datasets are publicly available at \url{https://github.com/shahinmomenzadeh/ALCMeans.git}.
	\end{abstract}
	

\section{Introduction}
Community Detection is a key topic in analyzing complex networks and one of the fundamental challenges in data science \cite{1}. This process identifies groups of nodes in a network where intra-group interactions are much stronger than inter-group interactions \cite{2}. Such communities often represent the underlying structures of a network, and understanding them helps to uncover hidden patterns, trends, and relationships in the data \cite{3}. Community detection has many applications in different domains. For example, social network analysis can identify groups of friends, users with common interests, or communities active on specific topics \cite{4}. Biological networks: In network biology, community detection can reveal functional pathways of proteins or genes, helping to understand diseases and drug development better \cite{5}. Financial networks: In economic systems, community analysis can identify groups of investors or institutions with similar behaviors. Transportation networks: Identifying clusters of stations or nodes with high traffic can help improve system planning and management \cite{6, 7}. There are several approaches to community detection in complex networks, each leveraging different principles to identify meaningful structures. Some methods are based on graph partitioning, which divides the network into subgroups by optimizing a global objective function, such as minimizing edge cuts \cite{8}. Others rely on statistical inference models, which assume that communities follow a probabilistic structure and estimate the most likely grouping of nodes \cite{9}. Another common approach is modularity optimization, which seeks to maximize the difference between observed intra-community links and expected connections in a random network \cite{10}. Among these methods, one of the most widely used techniques is clustering, which groups nodes based on similarity, proximity, or interaction patterns \cite{11}. Clustering is an effective approach because communities in complex networks typically exhibit dense intra-group connections and sparser inter-group interactions. By leveraging this property, clustering methods can reveal hidden structures without requiring explicit labels or predefined categories, making them well-suited for unsupervised learning scenarios. Furthermore, clustering is computationally more efficient than many global optimization-based methods \cite{12, 13}. However, existing approaches suffer from several critical limitations: The need for a predefined number of clusters: Many clustering algorithms require the number of communities to be specified in advance, which is impractical in real-world networks where the actual number of communities is unknown. A wrong estimation can lead to poor results \cite{14}. The accuracy of clustering strongly depends on the selection of representative nodes as cluster centers. Existing methods often struggle to effectively determine these centers, which leads to suboptimal community structures \cite{15}.

To overcome these challenges, this paper introduces Automatic Laplacian Centrality Means (ALCMean's), a novel algorithm designed to enhance the accuracy and efficiency of community detection. AlCMean's incorporates three key innovations:
\begin{enumerate}
	\item Automatic identification of cluster centers using local Laplacian energy, eliminating the need to predefine the number of communities and enabling more natural and reliable selection of representatives.
	\item DeepWalk-based representation learning, which captures richer structural information and ensures that nodes with similar embedding representations are grouped together.
	\item Proximity-based node assignment: Once cluster centers are identified, nodes are assigned based on their proximity to these centers, ensuring that each node belongs to its most relevant community without requiring predefined structures. This approach improves both adaptability and clustering accuracy.
\end{enumerate}

The key advantage of AlCMean's lies in its independence from prior assumptions and its robust performance across networks with diverse structures. Unlike traditional methods, AlCMean's integrates both structural centrality and representation learning, which allows it to consistently outperform baseline algorithms such as Louvain, Newman–Girvan, LPA, and Fast-Greedy, as well as more recent methods like MAGI \cite{16}. Experimental results demonstrate that AlCMean's achieves notable improvements on standard metrics such as NMI, ARI, modularity, and F1-score, validating its effectiveness in identifying well-structured communities. By addressing the core limitations of traditional clustering methods, AlCMean's introduces a powerful and innovative framework for analyzing complex networks across various domains. Nevertheless, its reliance on DeepWalk parameters suggests future research directions in adaptive embedding and scalable extensions.

The remainder of this paper is structured as follows: Section 2 reviews the related work. Section 3 presents the proposed AlCMean's algorithm in detail. Section 4 provides the experimental evaluation and comparisons with baseline methods. Section 5 discusses limitations and future directions, and Section 6 concludes the paper.
 
\section{Related Works}
Community detection in complex networks is one of the essential topics in network data analysis, with applications in social sciences, biology, and web analysis. It identifies groups of nodes in a network whose intra-group interactions are denser than their external interactions. Despite significant progress, challenges such as automatically detecting the number of communities, improving accuracy, and enhancing computational efficiency remain unsolved. This section reviews existing research on community detection and highlights recent advancements.

One notable method is K-rank, a variant of K-means designed to mitigate sensitivity to initial cluster centers \cite{17}. While traditional clustering methods like K-means are computationally efficient, they often degrade on large-scale networks and rely heavily on initialization quality. K-rank addresses these issues by introducing a rank-centrality–based initialization strategy and achieves faster convergence than K-means++. It also incorporates node similarity for classification and is applicable to directed, weighted, and overlapping networks. Experimental results on artificial and real networks demonstrate its effectiveness. However, K-rank still requires prior knowledge of the number of clusters and does not leverage representation learning, which limits its adaptability to noisy or sparse graphs. These limitations highlight the need for algorithms such as AlCMean’s, which combine structural centrality with embedding-based learning for greater robustness.

The following study introduces K-rank-D, an extension of K-rank that aims to identify communities without parameter tuning \cite{18}. This method leverages the topological structure of the network and selects influential, well-scattered nodes as initial centers, thereby addressing the limitations of K-means and K-rank in terms of initialization sensitivity. Experimental results show that K-rank-D achieves higher accuracy and efficiency in identifying communities and selecting centers compared to previous methods. Nevertheless, K-rank-D still inherits some drawbacks of density-based approaches, such as sensitivity to variations in community density and difficulty in handling heterogeneous graph structures. These issues restrict its generalization capability, reinforcing the need for more adaptive frameworks.

Another line of research introduces an algorithm for detecting overlapping communities in complex networks based on the Density Peaks method \cite{19}. Overlapping structures are particularly important in domains such as social and protein networks, where nodes may belong to multiple communities. This method constructs a similarity-based distance matrix and identifies community centers in a restrictive way, then assigns nodes according to a membership vector. Experimental results on synthetic and real networks confirm its effectiveness and better accuracy compared to traditional algorithms. However, the reliance on density thresholds and heuristic similarity measures makes the approach less robust in highly sparse networks or those with irregular community boundaries. This limitation underscores the need for alternative models, such as AlCMean’s, which avoid heuristic thresholds by combining Laplacian centrality with embedding-based assignment.

Another algorithm in this field is Laplacian centrality clustering (LPC), which differs from density-based delta clustering (DPC) by eliminating the need for empirical parameters \cite{20}. LPC transforms the data into a weighted graph and uses Laplacian centrality to evaluate the importance of each node, thereby identifying cluster centers automatically. Experimental results demonstrate that LPC outperforms DPC and other classical algorithms in clustering tasks. Nevertheless, its computational cost increases with graph size, and relying solely on Laplacian centrality may overlook higher-order structural features, limiting its scalability on large or feature-rich networks.

Another paper introduces the KNN-ADPC algorithm, which extends previous density peak clustering approaches by employing k-nearest neighbors to calculate distances and assign data points \cite{21}. Unlike classical DPC methods that require multiple sensitive parameters, KNN-ADPC needs only a single parameter. This makes it more efficient to configure and easier to apply in practice. Additionally, the algorithm uses an adaptive merging strategy. This strategy addresses the common problem of over-segmentation by ensuring excessively fragmented clusters are combined. As a result, the mechanism improves clustering accuracy and corrects misclassifications caused by local density variations. Experimental results show KNN-ADPC is effective for non-spherical and irregularly shaped data, a challenge for many traditional algorithms. However, it relies on a predefined neighborhood size, which can bias graphs with heterogeneous structures. Scalability to very large networks is also limited due to repeated nearest-neighbor searches.

Also, the paper reviews the DPC algorithm and its challenges \cite{22}. DPC has shown promise in various settings, but it encounters difficulties when communities have non-uniform densities or exist at different scales, often leading to fragmented or inaccurate results. To address these issues, an improved version called DPC-CE was proposed, which integrates graph connectivity evaluation to better simulate density and refine the identification of cluster centers. Experimental results indicate that DPC-CE performs more reliably than classical DPC, particularly in networks with non-spherical clusters and heterogeneous density distributions. However, both DPC and its improved variants remain sensitive to the choice of density thresholds and may struggle with scalability in very large networks, limiting their applicability in complex real-world scenarios.

Another contribution in this area is the Nonparametric Density Clustering (NAPC) algorithm, which was designed to overcome the limitations of conventional density-based methods such as DPC \cite{23}. NAPC introduces the use of divergence distance instead of Euclidean distance, allowing it to capture more nuanced differences in features and angular relationships between nodes. To further reduce dependency on sensitive parameters, NAPC leverages the Adjusted Boxplot theory to automatically adjust the density cutoff value. It also employs a novel index that combines local density with divergence distance for more accurate identification of cluster centers. Experimental evaluations on synthetic and real datasets show that NAPC achieves better clustering accuracy and robustness compared to classical density-based algorithms. Despite these improvements, the computational overhead introduced by divergence distance calculations can limit scalability, and the method may require careful adaptation when applied to very large or high-dimensional networks.

The LapEFCM method has also been introduced to detect overlapping communities in complex networks. Unlike traditional fuzzy clustering algorithms that require prior knowledge of the number of communities, LapEFCM eliminates this constraint by combining multiple steps. First, a local random walk algorithm is applied to compute a similarity matrix and extract node features. Next, the Laplacian maps technique reduces dimensionality, and fuzzy C-means clustering is used to identify overlapping communities. Experimental results on both real and artificial networks demonstrate that LapEFCM achieves better performance than comparable algorithms and can automatically determine the optimal number of communities. However, the multi-stage design increases computational complexity, and the reliance on fuzzy clustering may reduce stability when applied to highly dynamic or large-scale networks \cite{24}.

Another study investigates the problem of community deception in complex networks, where the goal is to deliberately hide specific communities from detection algorithms \cite{25}. The proposed method, called Commulet-AL, analyzes social and scientific networks to simulate the behavior of various community detection algorithms and to identify influential nodes within these communities. By strategically modifying network connections, Commulet-AL can reduce the detectability of targeted communities. While this approach provides valuable insights into the vulnerabilities of existing algorithms, it is primarily adversarial in nature and does not directly contribute to improving detection accuracy, highlighting an orthogonal line of research compared to conventional clustering methods.

More recently, researchers have revisited modularity maximization from the perspective of graph contrastive learning, leading to the development of the MAGI framework \cite{16}. MAGI formulates modularity maximization as a contrastive pretext task, using positive and negative pairs derived from the modularity matrix to guide representation learning. By combining modularity-based objectives with GNN encoders such as GCN and Graph SAGE, MAGI effectively mitigate semantic drift and eliminates the need for graph augmentations. Extensive experiments on small, large, and even extra-large networks (up to 100M nodes) show that MAGI consistently outperforms strong graph clustering baselines while remaining highly scalable. Nonetheless, MAGI’s reliance on GNN training introduces substantial computational overhead and requires careful hyperparameter tuning, which can limit its efficiency in resource-constrained scenarios.

In summary, a wide range of approaches have been developed for community detection in complex networks. Rank-based methods such as K-rank and K-rank-D improve initialization but still depend on parameter settings. Density-based techniques like DPC, DPC-CE, and NAPC enhance flexibility but remain sensitive to variations in density and scalability issues. Algorithms designed for overlapping communities, including Density Peaks–based methods and LapEFCM, improve modeling capability but often introduce additional computational costs. Centrality-driven clustering methods such as LPC automate center selection yet may overlook higher-order structural features. Recent advances in adversarial approaches such as Commulet-AL reveal vulnerabilities of detection methods but do not directly improve clustering accuracy. More recently, GNN-based frameworks like MAGI (KDD’24) have shown strong performance by integrating modularity maximization with contrastive learning, demonstrating both scalability and accuracy. However, across these diverse efforts, challenges remain in balancing accuracy, scalability, parameter independence, and stability. These open challenges motivate the need for novel approaches that integrate structural centrality with representation learning while avoiding heavy parameter tuning-directions that we pursue in this study.

\section{AlCMean's (Automatic Laplacian Centrality Means)}

This section introduces the design and implementation of the proposed AlCMean's algorithm. The method provides an innovative framework for community detection that eliminates the need to predefine the number of clusters, thereby addressing a major limitation of traditional approaches. The algorithm was implemented in Python 3.9 using well-established scientific libraries, including NetworkX for graph analysis, NumPy and SciPy for numerical computations, and Matplotlib/Seaborn for visualization.

The workflow of AlCMean’s consists of four main stages. First, a Laplacian energy score is computed for every node to evaluate its structural influence within the network. This score then sorts nodes, and the most influential ones initiate provisional communities together with their neighbors. Next, overlapping communities are merged to remove redundancy. For each resulting community, the node with the highest Laplacian energy is chosen as its center. Finally, DeepWalk embeddings are employed to capture higher-order structural information, and each node is assigned to the nearest community center in the embedding space. This multi-step procedure produces compact and meaningful community structures. The overall workflow is summarized in the following subsections, where each component of the algorithm is explained in detail.

\begin{algorithm}
	\caption{The AlCMean’s Algorithm}
	\textbf{Input:} Graph, vector size, number of walks, walk length \\
	\textbf{Output:} Detected communities
	\begin{algorithmic}[1]
		\State Compute Laplacian Energy for all nodes in G.
		\[
		E(V) = d(v)^2 + d(v) + 2 \sum_{u \in N_v} d(u)
		\]
		\State Sort nodes in descending order of \( E(v) \).
		\State Form seed clusters: each uncovered node \( v \) becomes a center and is grouped with its neighbors.
		\State Merge clusters iteratively while overlap \( S_{\text{max}} \geq 0.5 \).
		\State Select final cluster centers as nodes with maximum \( E(v) \) in each cluster.
		\State Generate node embeddings using Optimized DeepWalk.
		\State Apply K-Means on embeddings, initialized with the selected centers.
		\State Assign all nodes to the nearest center.
	\end{algorithmic}
	\textbf{Return:} Final Communities
\end{algorithm}

\begin{figure}[H]
	\centering
	\includegraphics[width=\textwidth]{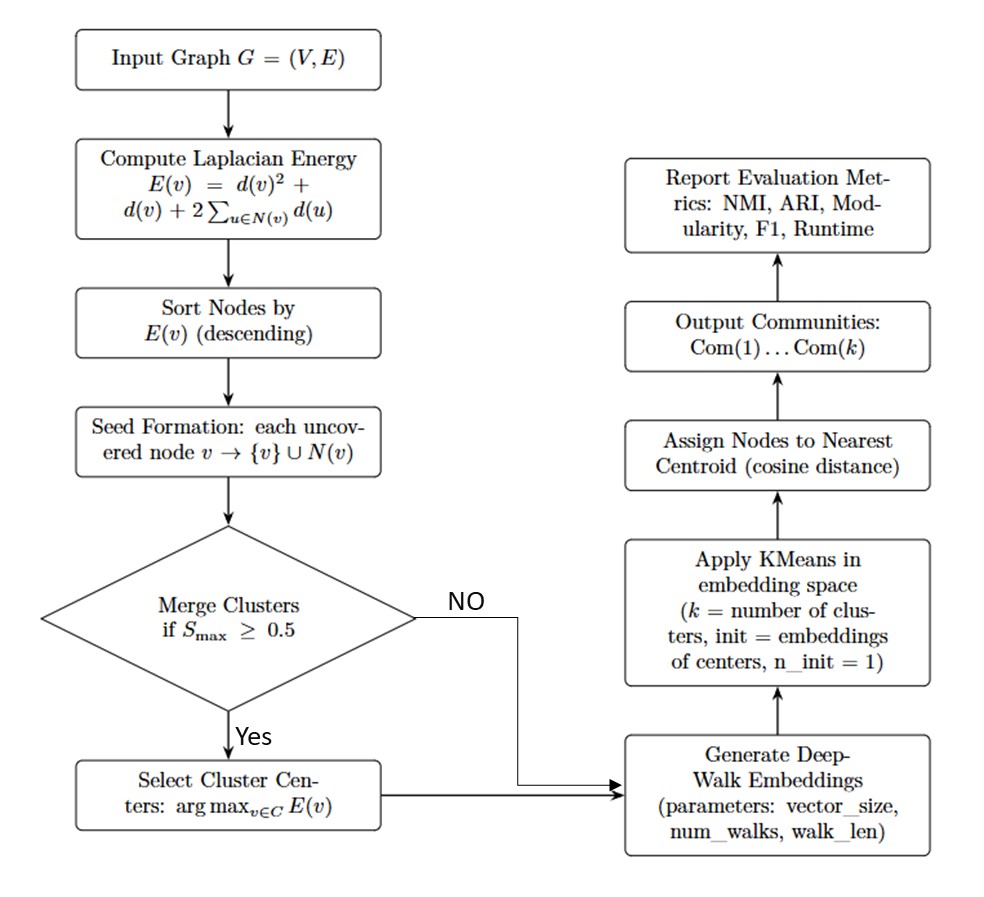}
	\caption{Flowchart of the AlCMean’s algorithm}
\end{figure}

\subsection{Node Importance via Laplacian Energy}

In the first stage of ALCMean’s, influential nodes are identified through a Laplacian energy score that reflects both their local connectivity and the structural context of their neighborhoods. This score combines the degree of each node with the degrees of its adjacent vertices, highlighting nodes that play a central role within their local substructures. It is important to note that ALCMean’s functions as a centralized, offline graph clustering algorithm. Consequently, all Laplacian energy values are computed directly from the global adjacency structure, and no scores are exchanged, transmitted, or propagated between nodes. Prior studies have demonstrated that Laplacian-based metrics reliably capture structural importance, making them well suited for identifying representative nodes in community detection tasks.

For a given node \( V \), the Laplacian energy is calculated as follows \cite{26}:
\[
E(V) = d(v)^2 + d(v) + 2 \sum_{u \in N_v} d(u)
\]

where \( d(v) \) is the degree of \( v \) and \( N(v) \) denotes its open neighborhood. Nodes with higher \( E(v) \) are considered more influential and are used to initialize provisional communities in the subsequent steps.

\subsection{Seed Formation}

After ranking all vertices by their Laplacian energy scores, the algorithm greedily initializes communities. The process iterates through nodes in descending order of \( E(v) \):
\begin{itemize}
	\item If a node \( v \) has not yet been covered by an existing cluster center, a new seed cluster is created as \( \{v\} \cup N(v) \), where \( N(v) \) denotes the neighborhood of \( v \).
	\item If \( v \) is already contained in a previously formed cluster, it is skipped to avoid redundancy.
\end{itemize}

This strategy ensures that high-energy vertices, which are structurally influential, serve as the anchors of initial communities. The inclusion of their neighbors guarantees local coverage around each center, while the restriction that seed centers must be distinct prevents overlap among the initial set of clusters. As a result, every node in the graph is either directly chosen as a center or is initially associated with a neighboring center, providing full coverage before the merging stage.

\subsection{Merging Communities}

The initial seed clusters often overlap because neighbors of different high-energy nodes may be shared. To refine the community structure, overlapping clusters are merged according to a similarity threshold. The similarity between two clusters \( C_i \) and \( C_j \) is defined asymmetrically as the fraction of shared nodes relative to the size of each cluster \cite{27}:
\[
\text{Similarity}(C_i) = \frac{|C_i \cap C_j|}{|C_i|}
\]
\[
\text{Similarity}(C_j) = \frac{|C_i \cap C_j|}{|C_j|}
\]
The maximum of these values is taken as the final similarity score:
\[
S_{\text{max}}(C_i, C_j) = \max (\text{Similarity}(C_i), \text{Similarity}(C_j))
\]
If \( S_{\text{max}}(C_i, C_j) \geq 0.5 \), the two clusters are merged. This process is repeated iteratively until no further merges are possible.

The choice of the 50\% threshold follows a majority-overlap heuristic:
\begin{itemize}
	\item Lower thresholds (e.g., 0.3) caused excessive merging, collapsing distinct communities into overly large clusters.
	\item Higher thresholds (e.g., 0.7) led to fragmented communities with significant redundancy.
\end{itemize}

Empirical validation across datasets showed that a threshold of 0.5 produced the most stable and interpretable results, and it is therefore adopted as the default setting.

\subsection{Selection of Cluster Centers}

After the merging stage, each community is represented by a single center node. The center of a community \( C \) is defined as the node with the maximum Laplacian energy score:
\[
\text{center}(C) = \arg \max_{v \in C} E(v)
\]
This rule ensures that the most structurally influential node within each cluster serves as its representative. Unlike average-based centroids, which may fall outside the discrete set of nodes, this medoid-style selection preserves the interpretability of centers as actual vertices in the network.

These selected centers play a dual role: (i) they anchor the local structure of each community, and (ii) they provide the initialization points for the subsequent embedding-based clustering stage. By relying on Laplacian energy rather than random initialization, AlCMean’s achieves more stable and reproducible results across different runs.

\subsection{Representation Learning with DeepWalk}

To capture higher-order structural information beyond local connectivity, AlCMean’s integrates DeepWalk into its pipeline. DeepWalk generates low-dimensional vector representations of nodes by simulating truncated random walks and applying the Word2Vec skip-gram model. This approach embeds nodes into a continuous feature space where proximity reflects structural similarity.

In our implementation, the following parameters are used:
\begin{itemize}
	\item Vector dimension (\( d \)): chosen from \( \{5, 7, 10\} \) depending on dataset size, balancing efficiency and embedding quality.
	\item Number of walks per node: varied between 50 and 90 to ensure adequate sampling of structural contexts.
	\item Walk length: selected between 20 and 60 steps to capture both local and moderately global patterns.
\end{itemize}

These ranges were validated through preliminary experiments and found to consistently produce robust embeddings across different datasets. By aligning the embedding stage with structurally influential centers identified earlier, AlCMean’s combines graph-theoretic importance with representation learning, thereby enhancing the stability and accuracy of community detection.

\subsection{Final Assignment}

In the final stage, community membership is refined using the learned embeddings. Each previously selected center is embedded via DeepWalk, and these vectors are used to initialize the clustering step. Specifically, K-Means is applied in the embedding space with:
\begin{itemize}
	\item The number of clusters is equal to the number of centers obtained after merging.
	\item The initial centroids are set to the embeddings of those centers.
\end{itemize}

This initialization strategy prevents randomness in centroid selection and ensures consistency across runs. The algorithm converges to stable partitions without repeated reinitialization. Each node \( v \in V \) is then assigned to the nearest centroid in the embedding space according to cosine distance. By combining Laplacian energy-based center identification with embedding-driven refinement, AlCMean’s avoids the pitfalls of arbitrary initialization and achieves both accuracy and reproducibility in community detection.

\subsection{Time Complexity Analysis}

\begin{itemize}
	\item Laplacian Energy Computation: \( O(|E|) \), where \( |E| \) is the number of edges, since degree information is derived from adjacency lists.
	\item Seed Formation and Sorting: Sorting nodes by energy requires \( O(|V| \log |V|) \), where \( |V| \) is the number of nodes.
	\item Merging Communities: In the worst case, this step requires \( O(|V|^2) \), but in practice it is much lower due to early merging convergence.
	\item Deep Walk Embedding: The dominant cost depends on the number of walks (\( \gamma \)), walk length (\( \ell \)), and dimension (\( d \)), yielding \( O(|V| \cdot \gamma \cdot \ell \cdot d) \).
	\item K-Means Clustering: With \( k \) clusters and \( t \) iterations, the complexity is \( O(|V| \cdot k \cdot t \cdot d) \).
\end{itemize}

Thus, the total complexity can be expressed as:
\[
O(|E| + |V| \log |V| + |V| \cdot \gamma \cdot \ell \cdot d + |V| \cdot k \cdot t \cdot d)
\]
Although the embedding stage adds overhead compared to purely heuristic methods, the complexity remains manageable and significantly lower than GNN-based alternatives.

\section{Evaluation}

The evaluation of AlCMean’s was carried out on five benchmark datasets commonly used in community detection research: Texas, Cornell, Washington, Football, and email-Eu-core. These datasets represent diverse domains and structural properties, ranging from social and academic networks to communication graphs. Their heterogeneity makes them well suited for testing both the accuracy and robustness of community detection algorithms.

Performance was measured using a set of complementary metrics. Normalized Mutual Information (NMI) and the Adjusted Rand Index (ARI) quantify the agreement between the detected communities and the ground truth partitions, with values closer to 1 indicating higher accuracy. Modularity (Q) evaluates the structural quality of the partition in terms of intra- versus inter-community connectivity. To further capture node-level classification quality, we report both F1-macro and F1-micro scores, which balance precision and recall across classes. Together, these metrics provide a comprehensive assessment of clustering effectiveness and community structure fidelity.

\begin{table}[H]
	\centering
	\caption{Data set in the proposed method}
	\vspace{3pt}
	\begin{tabular}{|c|c|c|c|}
		\hline
		\textbf{K} & \textbf{Edge} & \textbf{Node} & \textbf{Dataset} \\
		\hline
		12 & 611 & 115 & Football \\
		5 & 298 & 187 & Texas \\
		5 & 295 & 183 & Cornell \\
		5 & 446 & 230 & Washington \\
		42 & 25571 & 1005 & email-Eu-core \\
		\hline
	\end{tabular}
\end{table}

\subsection{Baseline Algorithms}

To provide a rigorous assessment of AlCMean’s, we compared its performance with both classical community detection heuristics and recent neural approaches. The classical baselines include:
\begin{itemize}
	\item Louvain: a widely used modularity optimization method known for scalability and competitive accuracy.
	\item Newman–Girvan: an edge-betweenness–based divisive algorithm, effective on small networks but computationally expensive.
	\item Label Propagation (LPA): an iterative method that assigns communities through majority voting among neighbors, offering low complexity but unstable results.
	\item Fast-Greedy: a modularity-based agglomerative approach that greedily merges nodes and clusters to maximize modularity.
\end{itemize}

In addition to these heuristics, we considered more recent models based on graph representation learning. Specifically, we included Graph SAGE and GAT, two representative graph neural networks (GNNs) that leverage neighborhood aggregation and attention mechanisms, respectively. Furthermore, we reported results for MAGI (KDD’24), a modularity-guided contrastive learning framework that integrates graph neural networks with community detection objectives.

This selection of baselines ensures that AlCMean’s is evaluated against both traditional algorithms and state-of-the-art learning-based methods, covering a broad spectrum of techniques in the community detection literature.

\subsection{Results and Comparisons}

The performance of AlCMean’s across the five benchmark datasets is summarized in Table 2, with comparisons to both classical algorithms and recent learning-based methods. The results demonstrate that AlCMean’s consistently achieves competitive or superior outcomes across the full set of evaluation metrics. On the Football network, which is characterized by well-defined communities, AlCMean’s reached the highest NMI (0.91) and ARI (0.89), surpassing both Louvain and LPA. This indicates that the combination of Laplacian-based center selection and embedding-driven refinement effectively captures the community structure with remarkable accuracy. The method also preserved modularity at a competitive level and achieved strong F1-macro and F1-micro scores, confirming its robustness at both global and node-level scales.

For Texas and Cornell, which are smaller academic networks with relatively sparse structures, AlCMean’s again produced the best NMI values (0.19 and 0.24, respectively). While modularity values were slightly lower than those of Louvain and Fast-Greedy, the higher NMI and ARI suggest that the detected clusters align more closely with the ground truth communities. This trade-off reflects the design of AlCMean’s, which prioritizes structural influence and representation learning over purely modularity-driven partitions.

On the Washington network, the results show modest improvements. AlCMean’s achieved NMI = 0.184 and ARI = 0.30, which were competitive with LPA and superior to MAGI in terms of community alignment. The modularity values remained strong, suggesting that even in relatively noisy graphs, the method maintains balance between structural cohesion and accuracy.

The email-Eu-core dataset, which represents a larger and denser communication graph, highlights the scalability of AlCMean’s. Here, the algorithm attained the best NMI (0.70) and ARI (0.39), outperforming MAGI and classical baselines. Although modularity was moderate, the F1-micro score of 0.48 indicates that the model captured community memberships at the node level more reliably than alternatives. Overall, these results confirm that AlCMean’s delivers strong and stable performance across diverse datasets. It outperforms traditional heuristics in terms of accuracy (NMI, ARI) while remaining competitive in modularity. Compared with recent neural baselines such as Graph SAGE, GAT, and MAGI, AlCMean’s avoids over-parameterization and achieves comparable or better accuracy with lower complexity in training.

\begin{table}[H]
	\centering
	\caption{(NMI, ARI, Modularity (Q), F1-macro, F1-micro) of different algorithms}
	\vspace{3pt}
	\begin{tabular}{|c|c|c|c|c|c|c|}
		\hline
		\textbf{Dataset} & \textbf{NMI} & \textbf{ARI} & \textbf{Q} & \textbf{F1-macro} & \textbf{F1-micro} & \textbf{Method} \\
		\hline
		email-Eu-core & 0.56 & 0.26 & 0.42 & 0.12 & 0.40 & Louvain \\
		email-Eu-core & 0.04 & 0.001 & 0.003 & 0.022 & 0.12 & Newman-Girvan \\
		email-Eu-core & 0.18 & 0.011 & 0.089 & 0.078 & 0.19 & LPA \\
		email-Eu-core & 0.44 & 0.16 & 0.37 & 0.09 & 0.32 & Fast-Greedy \\
		email-Eu-core & 0.42 & 0.16 & 0.13 & 0.20 & 0.27 & MAGI \\
		email-Eu-core & 0.70 & 0.39 & 0.22 & 0.19 & 0.48 & Proposed \\
		Football & 0.85 & 0.70 & 0.60 & 0.67 & 0.80 & Louvain \\
		Football & 0.35 & 0.14 & 0.40 & 0.05 & 0.21 & Newman-Girvan \\
		Football & 0.86 & 0.75 & 0.58 & 0.74 & 0.81 & LPA \\
		Football & 0.69 & 0.47 & 0.54 & 0.37 & 0.57 & Fast-Greedy \\
		Football & 0.74 & 0.55 & 0.38 & 0.66 & 0.69 & MAGI \\
		Football & 0.91 & 0.89 & 0.55 & 0.72 & 0.87 & Proposed \\
		Texas & 0.08 & 0.05 & 0.457 & 0.08 & 0.28 & Louvain \\
		Texas & 0.05 & 0.02 & 0.04 & 0.14 & 0.52 & Newman-Girvan \\
		Texas & 0.15 & 0.11 & 0.29 & 0.05 & 0.43 & LPA \\
		Texas & 0.09 & 0.08 & 0.451 & 0.08 & 0.32 & Fast-Greedy \\
		Texas & 0.10 & 0.02 & 0.20 & 0.27 & 0.36 & MAGI \\
		Texas & 0.19 & 0.07 & 0.459 & 0.03 & 0.232 & Proposed \\
		Cornell & 0.14 & 0.03 & 0.44 & 0.06 & 0.235 & Louvain \\
		Cornell & 0.09 & 0.04 & 0.26 & 0.14 & 0.42 & Newman-Girvan \\
		Cornell & 0.22 & 0.08 & 0.46 & 0.03 & 0.24 & LPA \\
		Cornell & 0.13 & 0.03 & 0.44 & 0.07 & 0.24 & Fast-Greedy \\
		Cornell & 0.08 & 0.04 & 0.29 & 0.29 & 0.33 & MAGI \\
		Cornell & 0.24 & 0.03 & 0.53 & 0.02 & 0.20 & Proposed \\
		Washington & 0.12 & 0.04 & 0.46 & 0.04 & 0.23 & Louvain \\
		Washington & 0.12 & 0.03 & 0.04 & 0.06 & 0.48 & Newman-Girvan \\
		Washington & 0.183 & 0.11 & 0.27 & 0.03 & 0.42 & LPA \\
		Washington & 0.11 & 0.04 & 0.55 & 0.04 & 0.25 & Fast-Greedy \\
		Washington & 0.10 & 0.06 & 0.23 & 0.25 & 0.31 & MAGI \\
		Washington & 0.184 & 0.3 & 0.47 & 0.03 & 0.20 & Proposed \\
		\hline
	\end{tabular}
	
\end{table}

\subsection{Ablation Study}

The ablation results reported in Table 3 highlight the importance of each component of the AlCMean’s framework. Across all datasets, the full model consistently achieved the highest overall performance, confirming the effectiveness of combining Laplacian energy, similarity-based merging, and DeepWalk embeddings. Removing Laplacian energy resulted in noticeable declines in accuracy. On the Football dataset, for example, ARI dropped from 0.891 in the full model to 0.670, and similar degradations were observed in Texas and Cornell. These results confirm that Laplacian energy is critical for selecting structurally meaningful centers. Disabling the merging step (\(S_{\text{max}} \geq 0.5\)) generally caused fragmentation and over-segmentation. The effect was particularly evident in Washington and Cornell, where modularity values declined from 0.477 and 0.536 in the full model to 0.396 and 0.450, respectively. This indicates that merging prevents the proliferation of redundant communities and stabilizes the detected structures. The most severe performance losses occurred when DeepWalk embeddings were removed. In all datasets, NMI and ARI dropped drastically, and in some cases, modularity even became negative (e.g., Washington and Cornell). For instance, in the email-Eu-core dataset, NMI plummeted from 0.705 to 0.300 and ARI from 0.390 to almost zero. Similarly, on Football, NMI fell from 0.919 to 0.168 and ARI from 0.891 to 0.011. These findings clearly show that embeddings are indispensable for capturing higher-order structural patterns and refining cluster assignments. Overall, the ablation study confirms that each module contributes to the robustness of AlCMean’s: Laplacian energy provides principled initialization, merging controls redundancy, and embeddings supply the representational depth required for accurate and stable community detection.

\begin{table}[H]
	\centering
	\caption{Ablation study results of AlCMean’s across benchmark datasets}
	\vspace{3pt}
	\begin{tabular}{|c|c|c|c|c|}
		\hline
		\textbf{Dataset} & \textbf{Variant} & \textbf{NMI} & \textbf{ARI} & \textbf{Modularity} \\
		\hline
		Texas & Full Model & 0.1960 & 0.0776 & 0.4595 \\
		& LapEnergy & 0.1613 & 0.0063 & 0.3592 \\
		& SmaxMerge & 0.1861 & 0.0665 & 0.3921 \\
		& DeepWalk & 0.1732 & 0.1098 & -0.0656 \\
		\hline
		email-Eu-core & Full Model & 0.7049 & 0.3905 & 0.2248 \\
		& LapEnergy & 0.6446 & 0.2060 & 0.1408 \\
		& SmaxMerge & 0.6649 & 0.1927 & 0.1188 \\
		& DeepWalk & 0.2995 & 0.0029 & 0.0361 \\
		\hline
		Washington & Full Model & 0.1841 & 0.0333 & 0.4772 \\
		& LapEnergy & 0.1766 & 0.0202 & 0.4224 \\
		& SmaxMerge & 0.1791 & 0.0244 & 0.3961 \\
		& DeepWalk & 0.1074 & 0.0277 & -0.0641 \\
		\hline
		Football & Full Model & 0.9192 & 0.8912 & 0.5583 \\
		& LapEnergy & 0.8608 & 0.6698 & 0.5174 \\
		& SmaxMerge & 0.8695 & 0.7423 & 0.4745 \\
		& DeepWalk & 0.1681 & 0.0114 & 0.0188 \\
		\hline
		Cornell & Full Model & 0.2407 & 0.0318 & 0.5357 \\
		& LapEnergy & 0.2239 & 0.0080 & 0.3874 \\
		& SmaxMerge & 0.2300 & 0.0239 & 0.4497 \\
		& DeepWalk & 0.0578 & -0.0173 & -0.0442 \\
		\hline
	\end{tabular}

\end{table}

\subsection{Discussion of Limitations}

While AlCMean’s demonstrates strong and stable performance across various benchmark datasets, several limitations should be acknowledged. First, the method depends on the DeepWalk embedding stage, which introduces additional computational cost compared to purely heuristic approaches. Although this cost is moderate relative to training graph neural networks, it can still become significant for very large-scale graphs.

Second, the performance of the algorithm is partially influenced by the choice of embedding parameters (vector size, number of walks, and walk length). As shown in Table 4, different datasets required slightly different parameter settings to achieve optimal results. This sensitivity indicates that some level of parameter tuning is necessary, which may limit out-of-the-box applicability.

Third, while the Laplacian energy criterion effectively identifies structurally important nodes, it primarily reflects degree-based properties. Networks with more complex dependency structures may benefit from incorporating additional centrality measures or multi-scale criteria.

Finally, although the current evaluation demonstrates that AlCMean’s outperforms or matches classical heuristics and remains competitive with recent GNN-based approaches, it does not yet exploit advanced neural architectures. Future work could explore the integration of more expressive embedding models or hybrid frameworks that combine the interpretability of graph-theoretic heuristics with the representational capacity of deep learning.

Another limitation of this study is the absence of formal statistical significance testing when comparing performance differences across methods. Although the empirical results show consistent improvements, incorporating statistical tests could strengthen the conclusions in future work.

Despite these limitations, the overall results suggest that AlCMean’s strikes a favorable balance between efficiency, accuracy, and scalability, making it a practical tool for community detection in diverse network settings.

\begin{table}[H]
	\centering
	\caption{Deep Walk parameters}
	\vspace{3pt}
	\begin{tabular}{|c|c|c|c|}
		\hline
		\textbf{Dataset} & \textbf{Vector size} & \textbf{Number of walks} & \textbf{Walk length} \\
		\hline
		Football & 7 & 90 & 60 \\
		email-Eu-core & 10 & 70 & 20 \\
		Texas & 5 & 90 & 60 \\
		Cornell & 7 & 90 & 40 \\
		Washington & 10 & 50 & 60 \\
		\hline
	\end{tabular}
	
\end{table}

The overall performance trends are further illustrated in Figure (2,3), which summarizes all evaluation metrics across methods for each dataset. The grouped bar charts clearly show that the proposed AlCMean’s achieves consistently higher accuracy in terms of NMI and ARI, particularly on the Football and email-Eu-core networks. Even in more challenging cases such as Texas and Cornell, the method demonstrates stable improvements over classical heuristics. In terms of Modularity, Louvain and Fast-Greedy occasionally achieve higher values, reflecting their optimization objective; however, AlCMean’s remains competitive while simultaneously delivering stronger F1-scores at both the macro and micro levels. These results highlight the ability of AlCMean’s to balance structural quality with alignment to ground truth labels, providing robust performance across networks of varying size and density.

\begin{figure}[H]
	\centering
	\includegraphics[width=\textwidth]{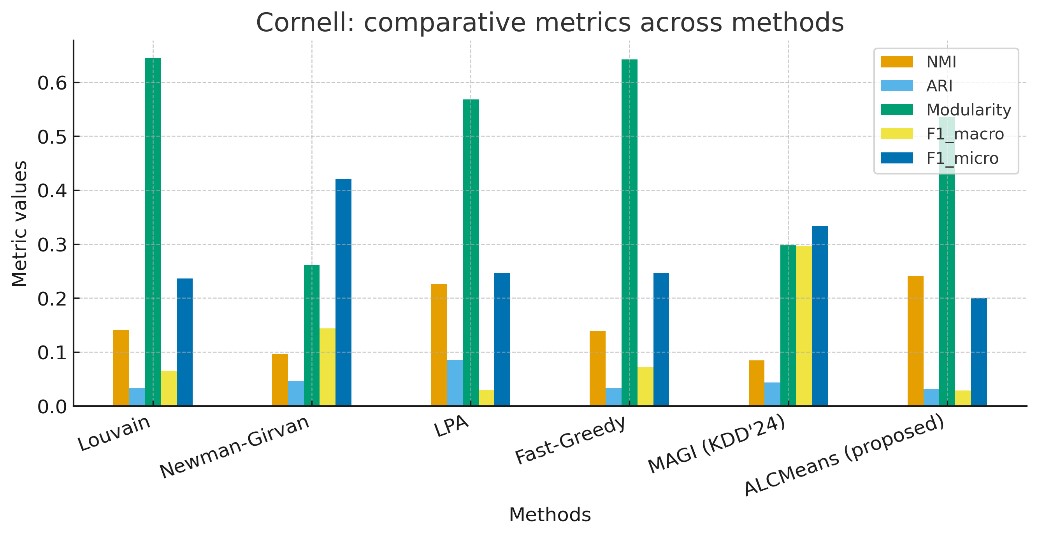}
	\includegraphics[width=\textwidth]{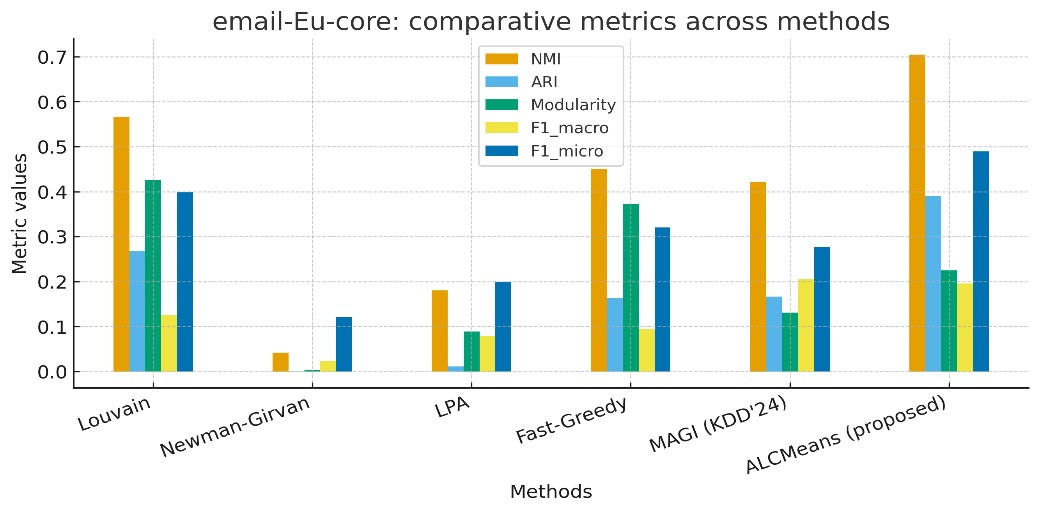}
	\caption{Comparison of performance metrics between AlCMean’s and other algorithms(cornell, email)}
\end{figure}

\begin{figure}[H]
	\centering
	\includegraphics[width=0.95\textwidth]{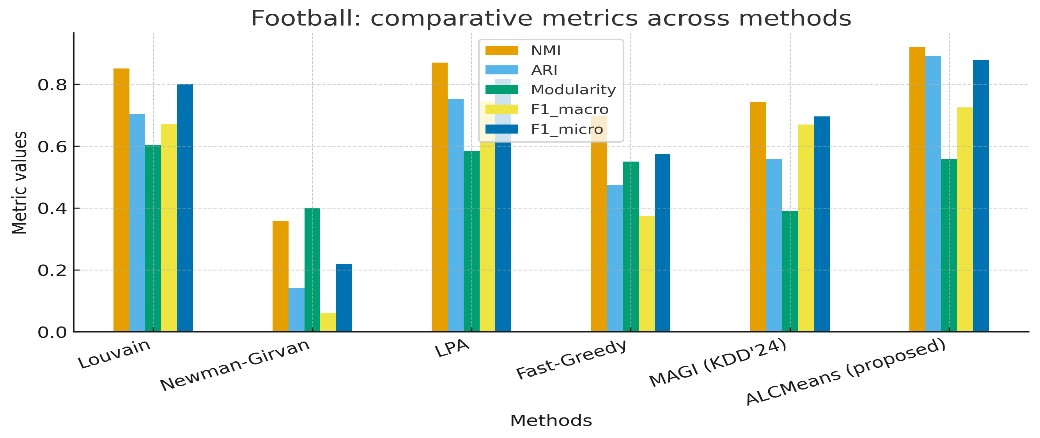}
	\includegraphics[width=0.95\textwidth]{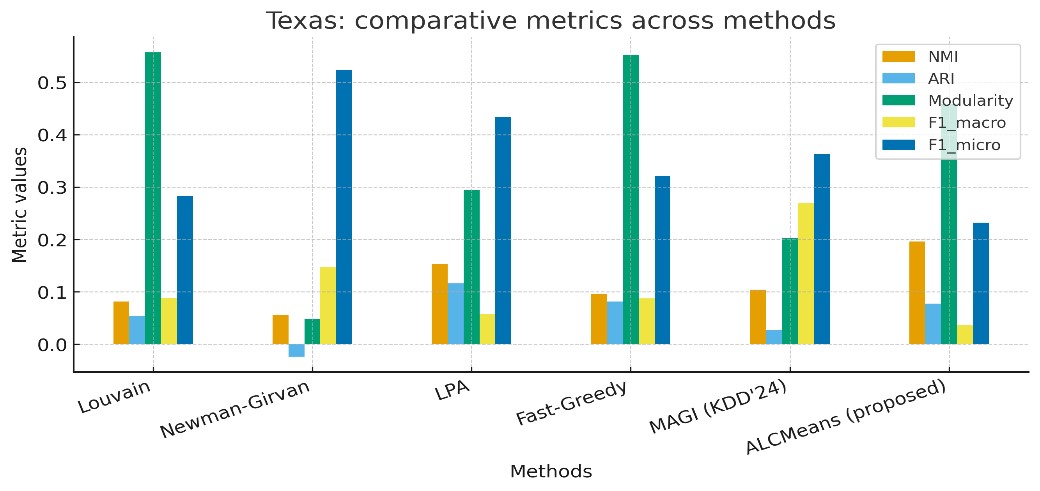}
	\includegraphics[width=0.95\textwidth]{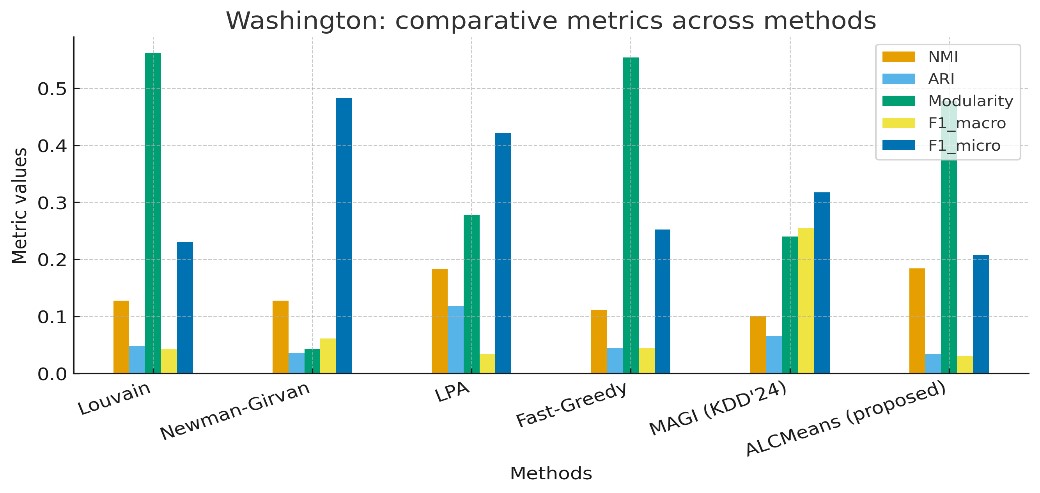}
	\caption{Comparison of performance metrics between AlCMean’s and other algorithms(football, taxas, washington)}
\end{figure}

\section{Conclusion}

This paper introduced AlCMean’s, a novel community detection algorithm that integrates Laplacian energy–based initialization, similarity-guided merging, and embedding-driven refinement. The method was designed to overcome key limitations of classical approaches, namely the need for a predefined number of clusters and sensitivity to random initialization. Extensive experiments across five benchmark datasets demonstrated that AlCMean’s consistently achieves high clustering accuracy, as measured by NMI and ARI, while maintaining competitive modularity and strong F1 performance. 

The ablation study confirmed that each component of the framework—Laplacian energy, merging, and DeepWalk embeddings—plays a critical role in the overall effectiveness of the method. Comparisons with both heuristic baselines and recent GNN-based approaches further showed that AlCMean’s offers a favorable trade-off between efficiency, accuracy, and scalability.

Despite its strengths, the algorithm has some limitations, including reliance on parameter tuning for the embedding stage and moderate computational cost compared to purely modularity-based heuristics. Future work could explore more adaptive embedding strategies, integration with advanced neural architectures, and extensions to dynamic or heterogeneous networks.

Overall, the results suggest that AlCMean’s is a robust and versatile framework for community detection, capable of delivering accurate and interpretable results across diverse network structures.

\end{document}